\documentclass[10pt,conference,letterpaper]{IEEEtran}
\usepackage{graphicx}
\usepackage{float}
\usepackage{amsmath}
\usepackage{amssymb}
\usepackage{enumerate}
\usepackage{booktabs}
\usepackage[table,xcdraw]{xcolor}
\begin{document}
\title{Control and Limit Enforcements for VSC Multi-Terminal HVDC in Newton Power Flow}
\author{
\IEEEauthorblockN{Hantao Cui, Fangxing Li}
\IEEEauthorblockA{Dept. of Electrical Engineering and Computer Science\\
University of Tenneseee\\
Knoxville, TN 37996\\
Email: \{hcui7, fli6\}@utk.edu}
\and
\IEEEauthorblockN{Haoyu Yuan}
\IEEEauthorblockN{Peak Reliability\\
	4850 Hahns Peak Dr., Suite 120\\
	Loveland, CO 80538 \\
	Email: hyuan@peakrc.com}

} %
\maketitle
\begin{abstract}
This paper proposes a novel method to automatically enforce controls and limits for Voltage Source Converter (VSC) based multi-terminal HVDC in the Newton power flow iteration process. A general VSC MT-HVDC model with primary PQ or PV control and secondary voltage control is formulated. Both the dependent and independent variables are included in the propose formulation so that the algebraic variables of the VSC MT-HVDC are adjusted simultaneously. The proposed method also maintains the number of equations and the dimension of the Jacobian matrix unchanged so that, when a limit is reached and a control is released, the Jacobian needs no re-factorization. Simulations on the IEEE 14-bus and Polish 9241-bus systems are performed to demonstrate the effectiveness of the method.
\end{abstract}

\section{Introduction}

The transmission network of the electric power system is undergoing a transformation with more renewable penetration through power converters. Voltage Sourced Converters (VSC), comparing to Silicon-Controlled Rectifiers, are advantageous with converter state control, independent active and reactive power control, and contingent power support capability. A number of VSC-based FACTS controllers such as STATCOM, SSSC and UPFC, have been deployed.

The imbalance between renewable generations and local power consumptions brings up the challenge of transmitting electricity continentally among asynchronous systems. Building a multi-terminal HVDC (MT-HVDC) overlay on the existing AC transmission grid is promising to improve transfer capacity and resilience of the grid, owing to VSC being able to make multi-terminal connections easily.

To understand the mechanism of VSC MT-HVDC and its impacts on the AC system, both the steady-state and transient process need to be modeled. Previous work has been carried out on the steady-state models of VSC HVDC such as UPFC \cite{nabavi1996steady}, IPFC \cite{zhang2006novel},  generalized models \cite{Zhang2004,baradar2013multi,gengyin2004power,milano2010power}, and generalized models with controls \cite{wang2014power,beerten2012generalized}. The dynamic models for transient analysis have been widely studied \cite{cole2010generalized,prieto2011methodology,beerten2014modeling}. Power flow analysis of systems with VSC is the basics for initializing the dynamic equations, however, the handling of controls limits are insufficient for software implementation.

Mathematically, the VSC controls versus the voltage and current limits is the choice of effective equations in the power flow analysis. In a typical AC power flow problem, when a PV generator is switch to a PQ load, a voltage angle variable is introduced along with a reactive power balancing equation. The same idea applies to network with VSC MT-HVDC, however, a typical implementation requires to change the size of the Jacobian matrix once a limit is violated.

In this paper, an automatic limit enforcement method for VSC MT-HVDC is proposed by including both the independent and dependent variables in the set. The number of equations and variables stays unchanged during the iteration process. The rest of this paper covers a generalized VSC MT-HVDC model with controls, incorporation of VSC MT-HVDC into Newton power flow, and the numerical results.

\section{Generalized VSC MT-HVDC Model}
\subsection{VSC MT-HVDC Equivalent Circuit}
\label{sec:VSC_circuit}
\begin{figure}
	\centering
	\includegraphics[trim={5cm 30cm 14.5cm 3.6cm},clip,scale=0.70]{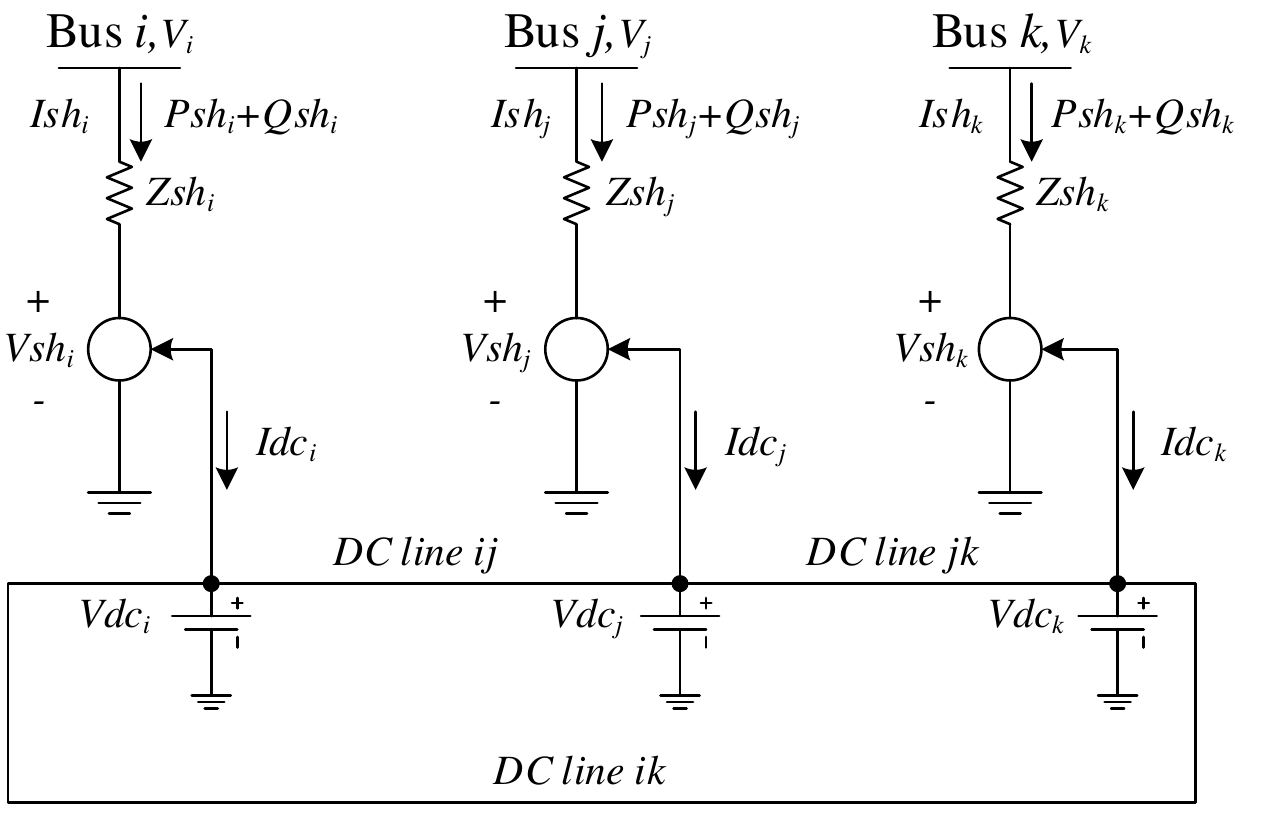}
	\caption{Generalized system with VSC MT-HVDC}
		\label{fig:generalized_mvsc_tdc_system_diagram}
\end{figure}

In power flow analysis, only the steady-state equations of VSC are considered. Fig. \ref{fig:generalized_mvsc_tdc_system_diagram} shows a VSC MT-HVDC equivalent circuit in shunt connection. The converters \textit{i, j, k} are connected to an AC network bus \textit{i, j, k } with a coupling transformer having an equivalent impedance of $Zsh$. They also connect to DC nodes \textit{i, j, k} linked by DC lines.

In this scheme, the converters at Buses \textit{i, j} are considered as primary converters which are capable of controlling the active power and reactive power flow from the AC buses independently. The converter at Bus \textit{k} is a secondary converter capable of controlling the AC bus voltage $V_k$ and the DC voltage $Vdc_k$. Therefore, converter \textit{k} is slack to balance power exchange among the converters. This scheme can be extended to N-terminal HVDC networks where the first $N-1$ converters are primary and the $N$th converter is secondary.

\subsection{VSC MT-HVDC Power Flow Equations}
\label{sec:VSC_power_flow}
The VSC MT-HVDC equivalent circuit is modeled in the phasor domain, i.e., the converters are represented at the fundamental frequency by the voltage phasors ${\bf{V}}s{h_m} = Vs{h_m}\angle \theta s{h_m} (m=i, j, k)$. The power injection from the AC bus $m$ to the coupling transformer is given by:

\begin{equation}
\label{eq:Sshm}
Ss{h_m} = V_m \times Ish_m^* = V_m \times {\left( {\frac{{{V_m} - Vs{h_m}}}{{Zs{h_m}}}} \right)^*}
\end{equation}
where $V_m$ is the voltage magnitude of bus $m$, $Ish_m$ is the current in the coupling transformer, and $Zsh_m$ the equivalent impedance of the transformer. $Zsh_m = Rsh_m + jXsh_m$, where $Rsh_m$ and $Xsh_m$ are the resistance and reactance of the transformer. The real and imaginary parts of \eqref{eq:Sshm} correspond to the active and reactive power injections:

\begin{equation}
\label{eq:Pshm}
\begin{split}
Psh_m = & gsh_m V_m^2 - gsh_m V_m Vsh_m\cos ({\theta _m} - \theta s{h_m}) \\
        & - bsh_m V_m Vsh_m\sin ({\theta _m} - \theta s{h_m})
\end{split}
\end{equation}

\begin{equation}
\label{eq:Qshm}
\begin{split}
Qsh_m = & -bsh_m V_m^2 - gsh_m V_m Vsh_m\sin ({\theta _m} - \theta s{h_m}) \\
        & + bsh_m V_m Vsh_m\cos ({\theta _m} - \theta s{h_m})
\end{split}
\end{equation}
where $Zsh_m = 1/(gsh_m + jbsh_m)$, $m=i, j, k$.

The power balancing equation of each converter involves the throughput power, converter losses and the actual output. The active power through the converter, $Pdc_m'$, is given by:

\begin{equation}
\label{eq:Pdcm'}
\begin{split}
Pdc_m' = & Re(-{\bf{V}}sh_m {\bf{I}} sh_m^*) \\
      = & gsh_m Vsh_m^2 - gsh_m V_m Vsh_m cos(\theta _m - \theta sh _m) \\
        & + bsh_m V_m Vsh_m sin(\theta _m - \theta sh_m) 
\end{split}
\end{equation}
which consists of two parts: the neat power injection to the DC network $Pdc_m$ and converter losses $Pl_m$. The converter loss term can be further split into three components: a constant power term, a constant voltage term, and a constant impedance term. In other words, the loss terms are independent, linearly and quadratically dependent on the converter current:
\begin{equation}
\label{eq:Pdcm_Pdcm'_Ploss}
\begin{split}
Pl_m & = Pdc_m' - Pdc_m \\
     & = a + b\cdot Ish_m + c\cdot Ish_m^2
\end{split}
\end{equation}

The next set of equations is the DC network equations. The DC network model composes of DC nodes and DC lines. The DC power flow pattern is dictated by the network line resistances and the DC node voltages following Kirchoff's laws. Therefore, a voltage variable and a current injection equation is added for each node.

The current injection from the DC network to the nodes  given in the following matrix form:
\begin{equation}
\label{eq:I_vector}
{\bf{I}} = -{\bf{Y}} {\bf{V}}dc
\end{equation}
where ${\bf{I}} = [I_i, I_j, I_k]^T$ is the DC nodal current injections from lines, ${\bf{V}}dc = [Vdc_i, Vdc_j, Vdc_k]^T$ is the DC node voltage magnitudes, and ${\bf{Y}}$ is the DC conductance matrix following the AC admittance matrix definition. Equation \eqref{eq:I_vector} can be written in tensorial form as:
\begin{equation}
\label{eq:Idc_m_tensorial}
Id{c_m} = -\sum\limits_n {{Y_{mn}} \cdot {V_n}} 
\end{equation}
where $\forall m=i,j,k$, $n = \{i, j, k\}$, and $Y_{mn}$ is the element at ($m, n$) in the conductance matrix $\bf{Y}$. 

The current injection from into node $m$ is given as:
\begin{equation}
\label{eq:Idc_m_injection}
{\bf{I}} dc_m = Pdc_m / V_m
\end{equation}
where $Pdc_m$ is the power injection from VSC into the DC network given by \eqref{eq:Pdcm_Pdcm'_Ploss}. The current injections from all devices into node $m$ follow Kirchoff's law and sum up to zero.

Finally, all converters are subject to physical voltage and current limits. The limits are rated values of the converters which need to be handled carefully in the power flow formulation. Voltage limits are given by:
\begin{equation}
\label{eq:V_m_limit}
Vsh_m^{min} \le Vsh_m \le Vsh_m^{max}
\end{equation}

The current flow through the VSC equals the current through the coupling transformer, given by:
\begin{equation}
\label{eq:Ish_expr}
Is{h_m} = \frac{{\sqrt {V_m^2 + Vsh_m^2 - 2{V_m}Vs{h_m}\cos ({\theta _m} - \theta s{h_m})} }}{{|Zs{h_m}|}}
\end{equation}

Note the throughput current is not an independent variable in the equations but a function of several variables. In the final solution, it is limited within the rating:
\begin{equation}
\label{eq:Ish_limit}
Ish_m \le Ish_m^{max}, m=i, j, k
\end{equation}

\subsection{Voltage and Power Flow Control of Converters}
In an N-terminal VSC HVDC network, the $N-1$ primary converters are capable of controlling either PQ or PV independently, while the $N$th secondary converter can control the voltage magnitude on the AC bus and DC node. Power or voltage control of VSC in power flow analysis forces the controlled variable at the desired value. These controls are valid if neither voltage or current constraint is binding. That is, the controlled variables are let equal the desired values, depending on the control mode.

\subsubsection{Primary VSC - PQ Control}
The primary converter controls the active and reactive power injections at the connected AC bus independently. This is given by:
\begin{equation}
\label{eq:PQ_P_control}
0 = Psh_m - Psh_m^c
\end{equation}
\begin{equation}
\label{eq:PQ_Q_control}
0 = Qsh_m - Qsh_m^c
\end{equation}
where $Psh_m^c$ and $Qsh_m^c$ are the desired active power and reactive power on bus $m$, $m=i,j$.

If either voltage or current limit of a PQ-controlled primary converter is reached, the voltage will be set to the limit to release the reactive power control. If the other limit is also reached afterwards, the active power control will be released.
\subsubsection{Primary VSC - PV Control}
The converter controls the active power injection and bus voltage magnitude on the AC bus. The control is given by \eqref{eq:PQ_P_control} and \eqref{eq:PV_V_control}:
\begin{equation}
\label{eq:PV_V_control}
0 = V_m - V_m^c
\end{equation}
where $V_m^c$ is the desired voltage magnitude of bus $m$, $m=i,j$.

If voltage or current limit is reached, the converter voltage will be set at the limit, and the AC voltage control will be dropped first, and then the active power control.

\subsubsection{Secondary VSC - Voltage Control}
The secondary VSC acts as a reference/slack bus for active power balancing in the DC network. DC nodal voltage on the DC slack node is controlled to the reference value given by:
\begin{equation}
\label{eq:V_Vdc_control}
0 = Vdc_k - Vdc_k^c
\end{equation}
Power injections on the connected AC bus is hence slack and uncontrollable. The voltage magnitude on the connected AC bus is controlled, given by:
\begin{equation}
\label{eq:V_V_control}
0 = V_k - V_k^c
\end{equation}
where $V_k^c$ is the desired voltage magnitude on the secondary VSC connected bus $k$. AC network voltage control will be released if either voltage limit or current limit is reached.

\subsection{Summary of VSC Power Flow Model}
\label{Model_Summary}
The generalized VSC MT-HVDC model with controls and limits contains the following equations:
\subsubsection{AC Bus Power Outputs} \eqref{eq:Pshm} and \eqref{eq:Qshm}
\subsubsection{VSC Power Injections and Losses} \eqref{eq:Pdcm'} and \eqref{eq:Pdcm_Pdcm'_Ploss} 
\subsubsection{VSC Limits} \eqref{eq:V_m_limit} and \eqref{eq:Ish_limit},
\subsubsection{Primary Converter Controls} \eqref{eq:PQ_P_control}, \eqref{eq:PQ_Q_control} or \eqref{eq:PQ_P_control}, \eqref{eq:PV_V_control}
\subsubsection{Secondary Converter Controls} \eqref{eq:V_Vdc_control}, \eqref{eq:V_V_control}
\subsubsection{DC Network Current Injections} \eqref{eq:Idc_m_injection}

\section{Formulating into Newton Power Flow}
The power flow problem is to find the zero of a set of non-linear equations starting from an adequate initial guess. The general form of the power flow equations is given as follows:
\begin{equation}
\label{eq:Newton_General}
{\bf{g(y)}} = 0
\end{equation}
where {\bf{y}} is the steady-state algebraic variables. The state variables of the differential equations will be initialized afterwards.

\subsection{Newton Method with Automatic Reactive Power Limit}
\label{sec:Newton_method_auto_limit}
Before considering VSC MT-HVDC, the power flow equations are revisited. The commonly adopted equations are nodal power mismatches which include the active power mismatch for PQ- and PV-connected buses, and reactive power mismatch for PQ-connected buses. Reactive power limits of PV generators are checked after each iteration. If a limit is reached, the PV generator will be converted to a PQ load to fix the reactive power output at the limit, and a reactive mismatch equation has to be added. Checking the reactive power limit can be easily done by a if-then logic, however, the addition of equations will change the size of matrices and requires re-factorization of the Jacobian, which is time consuming.

Inclusion of PV reactive power output in the power flow equation set is proposed in \cite{milano2010power} which retains the size of the Jacobian matrix. For each PV-connected bus, a variable $Q$ for the reactive power output is added, so is an equation for the reactive power mismatch on that bus. Also added is a variable $V$ for the voltage magnitude and an equation for the voltage mismatch given by \eqref{eq:V_mismatch}, where $V_0$ is the desired value.

\begin{equation}
\label{eq:V_mismatch}
0 = V - V_0
\end{equation}

Organize the equations by grouping together active power mismatches, reactive power mismatches, voltage deviation (for PV and slack buses), and angle deviation (for the slack bus only), the equations can be written in the following form:

\begin{equation}
\label{eq:g_equation}
{\bf{g}} = [g_p^T, g_q^T, g_v^T, g_\theta ^T]^T = 0
\end{equation}
The linearized equation for each Newton iteration can be written as:
\begin{equation}
\label{eq:Iter_form}
\left[ {\begin{array}{*{20}{c}}
	{\Delta P}\\
	{\Delta Q}\\
	{\Delta V_e}\\
	{\Delta \theta_e }
	\end{array}} \right] =  - \left[ {\begin{array}{*{20}{c}}
	{\left. {\underline {\, 
				{\begin{array}{*{20}{c}}
					{{g_{p,\theta }}}&{{g_{p,v}}}\\
					{{g_{q,\theta }}}&{{g_{q,v}}}
					\end{array}} \,}}\! \right| }&{\begin{array}{*{20}{c}}
		0&{{g_{p,pg}}}\\
		{{g_{q,qg}}}&0
		\end{array}}\\
	{\begin{array}{*{20}{c}}
		0&{{g_{v,v}}}\\
		{{g_{\theta ,\theta }}}&0
		\end{array}}&{\begin{array}{*{20}{c}}
		\varepsilon &0\\
		0&\varepsilon 
		\end{array}}
	\end{array}} \right]\left[ {\begin{array}{*{20}{c}}
	{\Delta \theta }\\
	{\Delta V}\\
	{{Q_g}}\\
	{{P_g}}
	\end{array}} \right]
\end{equation}
where ${g_{p,\theta }} = \nabla ^T_\theta {g_p}$, ${g_{p,v }} = \nabla ^T_v {g_p}$,  ${g_{q,\theta }} = \nabla ^T_\theta {g_q}$, ${g_{q,v }} = \nabla ^T_v {g_q}$, ${g_{v,v }} = \nabla ^T_v {g_v}$, ${g_{\theta,\theta }} = \nabla ^T_\theta {g_\theta}$. Note that ${g_{p,pg }} = \nabla ^T_{pg} {g_p}$ and ${g_{q,qg }} = \nabla ^T_{qg} {g_q}$ are the derivatives of $g_p$ and $g_q$ with respect to the specific generator output, and $\epsilon$ is a diagonal matrix of small values ($10^{-6}$) to avoid singularity in matrix factorization.

Each time before evaluating all the equations, the reactive power limits are checked for violations. If violation happens, the corresponding reactive power output will be set to the limit, and, more importantly, the voltage mismatch equations will be forced at 0, which invalidate the voltage control on PV buses. This process does not affect the size or shape of the Jacobian matrix, hence the symbolic factorization can be re-used.

One observation from the Jacobian matrix \eqref{eq:Iter_form} is that, except for the upper-left block corresponding to the bus power injection mismatch equations, the values in matrix are constant. In other words, only the upper-left block needs to be updated at every iteration using the Newton method. Since the matrix size does not change, some variations of Newton method which does not update the Jacobian at every step, e.g. the Dishonest Newton method can be applied.
 
\subsection{Incorporation of VSC MT-HVDC Model}
The equations and the Jacobian matrix need to be extended to incorporate the VSC MT-HVDC model described in Section \ref{Model_Summary}. The introduced variable set of VSC are given by \eqref{eq:VSC_vars}, whose increments are appended to the right-hand side of \eqref{eq:Iter_form}.
\begin{equation}
\label{eq:VSC_vars}
{\bf{X_2}} = [\theta sh,Vsh,Psh,Qsh,Pdc',Pdc,Vdc,Ish]
\end{equation}

The introduced equation set of VSC is given by \eqref{eq:g_VSC}, which are appended to the left-hand side of \eqref{eq:Iter_form}. The components in \eqref{eq:g_VSC} corresponds to equations \eqref{eq:Pshm}, \eqref{eq:Qshm}, \eqref{eq:PQ_P_control}, \eqref{eq:PQ_Q_control}, \eqref{eq:Pdcm'}, \eqref{eq:Pdcm_Pdcm'_Ploss}, \eqref{eq:Idc_m_injection}, \eqref{eq:Ish_expr}, respectively. All terms in each equation are moved to one side to evaluate the mismatch for each iteration.
\begin{equation}
\label{eq:g_VSC}
\begin{split}
{\bf{g_2}} = [ {g_{Psh}^T}, & {g_{Qsh}^T},{g_{Ps{h^c}}^T},{g_{Qs{h^c}}^T}, {g_{Pdc'}^T},{g_{Pl}^T},{g_{Pdc}^T},{g_{Ish}^T}]^T
\end{split}
\end{equation}

In addition to \eqref{eq:VSC_vars} and \eqref{eq:g_VSC}, for each VSC in voltage control mode, \eqref{eq:PV_V_control} or \eqref{eq:V_V_control} is not explicitly used. Rather, the connected AC bus is converted to a PV-type bus, where a voltage variable and voltage mismatch equation $g_{V^c}$ is added like \eqref{eq:V_mismatch}.

The corresponding Jacobian matrix is obtain by taking the derivative of each equation with respect to each variable. Note that the power outflos from the VSC connected AC buses are added to the AC network equations, namely $\Delta P$ and $\Delta Q$, the derivatives of $\Delta P$ and $\Delta Q$ with respect to $Psh$ and $Qsh$ needs to be evaluated. Similarly, the derivatives of \eqref{eq:g_VSC} with respect to AC voltage magnitude and angle need to be included.

\subsection{Automatic VSC Control and Limit Enforcing}

In the framework of Newton method with automatic reactive power limit described in Section \ref{sec:Newton_method_auto_limit}, the voltage and current limits of VSC MT-HVDC model can be automatically handled without changing the size of Jacobian matrix. The approach is described as follows:

\subsubsection{No Limit Violation}
If there is no limit violation at this iteration, all controls are maintained, which means enforcing the corresponding control equation. For example, if the PQ control of a primary converter is effective, $g_{Psh^c}$ and $g_{Qsh^c}$ need to be evaluated for mismatches.

\subsubsection{Limit Violations}
If either voltage or current limit is reached, the reactive power control or voltage control is first released. The violated term is set to the limit value, and the corresponding equation $g_{Qsh^c}$ or $g_{V^c}$ is set to 0 using \eqref{eq:g_release_control}. If the other limit is reached after releasing the first controlled variable, the active power control will be released. For the secondary converter, AC voltage control is released first, and then the DC node voltage control.

\begin{equation}
\label{eq:g_release_control}
{g_{Qs{h^c}}} \buildrel \Delta \over = 0\;or\;{g_{{V^c}}} \buildrel \Delta \over = 0
\end{equation}

The voltage and current limits of VSC can be checked at every iteration, but it is more effective to start enforcing the limits when the mismatch is relative small. 

\section{Case Studies and Results}

The proposed VSC MT-HVDC control and limit enforcement method is simulated on IEEE 14-bus and Polish 9241-bus systems. Simulations are performed on a Python-based software package, \textit{Andes} \cite{Cui2016}, using CVXOPT 1.1.8 for sparse matrix operations and KLU for fast sparse matrix factorizations. A generic Newton-Raphson method with a convergence tolerance of $10^{-8}$. Limit enforcement is enabled since the fourth iteration, and a maximum of one PV can be converted in each iteration. All the case studies are carried out on a laptop computer with a i5-6200U processor and 8GB RAM.

On the IEEE 14-bus system, we consider three scenarios:

\begin{enumerate}
	\item Base case with $qG_2 \le 0.4$.
	\item Base case with $qG_2 \le 0.4$, DC networks, and VSC
	\item Base case with $qG_2 \le 0.4$, DC networks, and VSC with reduced $I_{sh}$ limit on VSC 3.
		
\end{enumerate}

In the modified test case, generator reactive power is limited to 0.4 pu, 0.4 pu, 0.24 pu and 0.24 pu, respectively. The four VSCs are connected to the 14-bus system on Buses 1, 3, 12, and 14, and their DC output is connected to a circular DC network where each DC line has a resistance of 1 pu. The control methods and the parameters are listed in Table \ref{tab:hvdc_control_params}. All the loss coefficients are neglected.

\begin{table}[]
	\centering
	\caption{VSC MTDC system data}
	\label{tab:hvdc_control_params}
	\begin{tabular}{@{}lllllllll@{}}
		\toprule
		\begin{tabular}[c]{@{}l@{}}VSC\\ n\end{tabular} & \begin{tabular}[c]{@{}l@{}}Bus\\ m\end{tabular} & Ctrl 1 & \begin{tabular}[c]{@{}l@{}}Ctrl 1\\ Value\end{tabular} & Ctrl 2 & \begin{tabular}[c]{@{}l@{}}Ctrl 2\\ Value\end{tabular} & \begin{tabular}[c]{@{}l@{}}Vsh\\ max\end{tabular} & \begin{tabular}[c]{@{}l@{}}Vsh\\ min\end{tabular} & \begin{tabular}[c]{@{}l@{}}Ish\\ max\end{tabular} \\ \midrule
		1                                               & 1                                               & P      & 0.2                                                    & Q      & 0.1                                                    & 1.0                                               & 0.95                                              & 1                                                 \\
		2                                               & 3                                               & P      & 0.2                                                    & Q      & 0.1                                                    & 1.08                                              & 1.02                                              & 1                                                 \\
		3                                               & 12                                               & P      & 0.2                                                    & Vm     & 1.05                                                    & 1.1                                               & 0.9                                               & 1                                                 \\
		4                                               & 14                                              & Vm     & 1.035                                                  & Vdc    & 1.0                                                    & 1.1                                               & 0.9                                               & 1                                                 \\ \bottomrule
	\end{tabular}
\end{table}

 The base case solution to the original 14-bus system can be found in \cite{milano2010power}. The solution to the first scenario is listed in Table \ref{tab:scenario_1_results}, where reactive violation on bus 2 is enforced at the fourth iteration to fix the reactive power generation at its maximum. It takes 7 iterations in 0.0069 second to reach the tolerance.

\begin{table}[]
	\centering
	\caption{Scenario 1: Base case with $qG_2 \le 0.4$}
	\label{tab:scenario_1_results}
	\begin{tabular}{@{}lllllll@{}}
		\toprule
		\begin{tabular}[c]{@{}l@{}}$Bus$\\ m\end{tabular} & \begin{tabular}[c]{@{}l@{}}$v$\\ {[}pu{]}\end{tabular} & \begin{tabular}[c]{@{}l@{}}$\theta$\\ {[}rad{]}\end{tabular} & \begin{tabular}[c]{@{}l@{}}$p_G$\\ {[}pu{]}\end{tabular} & \begin{tabular}[c]{@{}l@{}}$q_G$\\ {[}pu{]}\end{tabular} & \begin{tabular}[c]{@{}l@{}}$p_L$\\ {[}pu{]}\end{tabular} & \begin{tabular}[c]{@{}l@{}}$q_L$\\ {[}pu{]}\end{tabular} \\ \midrule
		1                                                 & 1.06                                                   & 0                                                            & 2.3239                                                  & -0.1535                                                 & 0                                                       & 0                                                       \\
		2                                                 & 1.0442                                                 & -0.0015                                                     & 0.4                                                     & 0.4                                                     & 0.217                                                   & 0.127                                                   \\
		3                                                 & 1.01                                                   & -0.0039                                                     & 0                                                       & 0.2400                                                 & 0.942                                                   & 0.19                                                    \\
		4                                                 & 1.0183                                                 & -0.0032                                                     & 0                                                       & 0                                                       & 0.478                                                   & -0.039                                                  \\
		5                                                 & 1.0199                                                 & -0.0027                                                     & 0                                                       & 0                                                       & 0.076                                                   & 0.016                                                   \\
		6                                                 & 1.07                                                   & -0.0043                                                     & 0                                                       & 0.1243                                                 & 0.112                                                   & 0.075                                                   \\
		7                                                 & 1.0618                                                 & -0.0041                                                     & 0                                                       & 0                                                       & 0                                                       & 0                                                       \\
		8                                                 & 1.09                                                   & -0.0041                                                     & 0                                                       & 0.1744                                                 & 0                                                       & 0                                                       \\
		9                                                 & 1.0562                                                 & -0.0046                                                     & 0                                                       & 0                                                       & 0.295                                                   & -0.046                                                \\
		10                                                & 1.0512                                                 & -0.0046                                                      & 0                                                       & 0                                                       & 0.09                                                    & 0.058                                                   \\
		11                                                & 1.057                                                  & -0.0045                                                     & 0                                                       & 0                                                       & 0.035                                                   & 0.018                                                   \\
		12                                                & 1.0552                                                 & -0.0046                                                     & 0                                                       & 0                                                       & 0.061                                                   & 0.016                                                   \\
		13                                                & 1.0504                                                 & -0.0046                                                     & 0                                                       & 0                                                       & 0.135                                                   & 0.058                                                   \\
		14                                                & 1.0357                                                 & -0.0049                                                     & 0                                                       & 0                                                       & 0.149                                                   & 0.05                                                    \\ \bottomrule
	\end{tabular}
\end{table}

Scenario 2 considers the four VSC MT-HVDC and a DC network. The bus-wise solution to this scenario is listed in Table \ref{tab:scenario2-bus-results}, while the results of the VSC converters are given in Table \ref{tab:scenario2-vsc-results}, where the grey cells are the effective limits. Note the power flow into of VSC, namely $P_{sh}$ and $Q_{sh}$, are counted into the load on the connected buses. 

Three limits are violated at the fourth iteration: $q_{G2}$, $Vshmax_1$ and $Vshmin_2$, therefore, the reactive power control on VSC 1 and VSC 2 are dropped. On the contrary, as the iteration continues, $q_{G2}$ returned within its limit and the voltage on Bus 3 remained at 1.1 pu. The solution process takes 14 iterations in 0.0571 second to finish. For comparison, a case with wide ranges of voltage and higher current limits converges in 9 iterations in 0.015 second. Obviously, handling of the violations increases iterations and the calculation time.

\begin{table}[]
	\centering
	\caption{Scenario 2: Base case with $q_{G2}\le0.4$, VSC MT-HVDC network}
	\label{tab:scenario2-bus-results}
	\begin{tabular}{@{}lllllll@{}}
		\toprule
		\begin{tabular}[c]{@{}l@{}}$Bus$\\ m\end{tabular} & \begin{tabular}[c]{@{}l@{}}$v$\\ {[}pu{]}\end{tabular} & \begin{tabular}[c]{@{}l@{}}$\theta$\\ {[}rad{]}\end{tabular} & \begin{tabular}[c]{@{}l@{}}$pG$\\ {[}pu{]}\end{tabular} & \begin{tabular}[c]{@{}l@{}}$qG$\\ {[}pu{]}\end{tabular} & \begin{tabular}[c]{@{}l@{}}$pL$\\ {[}pu{]}\end{tabular} & \begin{tabular}[c]{@{}l@{}}$qL$\\ {[}pu{]}\end{tabular} \\ \midrule
		1                                                 & 1.06                                                   & 0                                                            & 2.3044                                                  & 0.4664                                                 & 0.2                                                     & 0.6169                                                  \\
		2                                                 & 1.045                                                  & -0.0799                                                     & 0.4                                                     & 0.3620                                                 & 0.217                                                   & 0.127                                                   \\
		3                                                 & 1.01                                                   & -0.226                                                       & 0                                                       & 0.1764                                                 & 1.142                                                   & 0.0712                                                 \\
		4                                                 & 1.023                                                  & -0.1575                                                     & 0                                                       & 0                                                       & 0.478                                                   & -0.039                                                  \\
		5                                                 & 1.0246                                                 & -0.1331                                                     & 0                                                       & 0                                                       & 0.076                                                   & 0.016                                                   \\
		6                                                 & 1.07                                                   & -0.2026                                                      & 0                                                       & 0.1799                                                 & 0.112                                                   & 0.075                                                   \\
		7                                                 & 1.0591                                                 & -0.1757                                                     & 0                                                       & 0                                                       & 0                                                       & 0                                                       \\
		8                                                 & 1.09                                                   & -0.1757                                                     & 0                                                       & 0.1915                                                 & 0                                                       & 0                                                       \\
		9                                                 & 1.0467                                                 & -0.1852                                                     & 0                                                       & 0                                                       & 0.295                                                   & -0.0422                                                \\
		10                                                & 1.0436                                                 & -0.1932                                                     & 0                                                       & 0                                                       & 0.09                                                    & 0.058                                                   \\
		11                                                & 1.0534                                                 & -0.1999                                                     & 0                                                       & 0                                                       & 0.035                                                   & 0.018                                                   \\
		12                                                & 1.05                                                   & -0.2382                                                     & 0                                                       & 0                                                       & 0.261                                                   & -0.1025                                                \\
		13                                                & 1.0524                                                 & -0.1983                                                     & 0                                                       & 0                                                       & 0.135                                                   & 0.058                                                   \\
		14                                                & 1.035                                                  & -0.1027                                                      & 0                                                       & 0                                                       & -0.4844                                                & 0.3010                                                 \\ \bottomrule
	\end{tabular}
\end{table}

\begin{table}[]
	\centering
	\caption{Scenario 2: VSC MT-HVDC power, voltage and current results}
	\label{tab:scenario2-vsc-results}
	\begin{tabular}{@{}llllllll@{}}
		\toprule
		\begin{tabular}[c]{@{}l@{}}Bus\\ $m$\end{tabular} & \begin{tabular}[c]{@{}l@{}}Node \\ $n$\end{tabular} & \begin{tabular}[c]{@{}l@{}}$Psh$\\ {[}pu{]}\end{tabular} & \begin{tabular}[c]{@{}l@{}}$Qsh$\\ {[}pu{]}\end{tabular} & \begin{tabular}[c]{@{}l@{}}$Pdc$\\ {[}pu{]}\end{tabular} & \begin{tabular}[c]{@{}l@{}}$Vm$\\ {[}pu{]}\end{tabular} & \begin{tabular}[c]{@{}l@{}}$Vsh$\\ {[}pu{]}\end{tabular} & \begin{tabular}[c]{@{}l@{}}$Ish$\\ {[}pu{]}\end{tabular} \\ \midrule
		1                                                 & 1                                                   & \cellcolor[HTML]{C0C0C0}0.2                              & 0.6169                                                   & -1.1811                                                 & 1.06                                                    & \cellcolor[HTML]{C0C0C0}1                                & 0.6118                                                   \\
		3                                                 & 2                                                   & \cellcolor[HTML]{C0C0C0}0.2                              & -0.1188                                                 & -0.2345                                                 & 1.01                                                    & \cellcolor[HTML]{C0C0C0}1.02                             & 0.2303                                                  \\
		12                                                & 3                                                   & \cellcolor[HTML]{C0C0C0}0.2                              & -0.1849                                                  & -0.2344                                                  & \cellcolor[HTML]{C0C0C0}1.05                            & 1.0596                                                   & 0.2214                                                   \\
		14                                                & 4                                                   & -0.6333                                                 & 0.25102                                                  & 0.6524                                                  & \cellcolor[HTML]{C0C0C0}1.035                           & 1.0189                                                   & 0.6583                                                  \\ \bottomrule
	\end{tabular}
\end{table}

Scenario 3 studies the impact of $Ish$ limit of VSC 3 on the power flow and voltage on Bus 12. In this scenario, we enforce $Ish_3 \le 0.19$ pu, which is smaller than the solution in Table \ref{tab:scenario2-vsc-results}. As a consequence, the VSC current limit will be violated and the voltage control on Bus 12 will be dropped. The power flow solution takes 29 iterations in 0.051 second to finish, and the results are listed in Table \ref{tab:scenario3_results_VSC}. The limit violations are reported as follows:

\begin{enumerate}
	\item At iteration 4, $Vshmax_1$, $Vshmin_2$ and $Ishmax_3$ are violated. $q_{G2}$ is also violated. 
	\item At iteration 6, $q_{G3}$ is violated. 
	\item At iteration 23, $Vshmax_3$ is violated.
\end{enumerate} 

This is an extreme case where both the voltage and active power controls on VSC 3 are dropped. The active power drawn on Bus 12 are sharply reduced to meet the current limit. Due to the power balancing equations in the DC network, if $Ishmax_3$ is too small, for example, $Ishmax_3\le0.18$, the power flow iteration will not converge. This happens when all the primary VSC active power control are valid, while the secondary cannot maintain the power balance in the DC network without violating its limit.

The Polish 9248-bus test system from MATPOWER is also studied. The base 9248-bus system is first solved, and a modified system with a 10-terminal VSC HVDC network is solved. PV reactive power checking for the system is turned off, while the limit enforcements of VSC remain on. The base case is solved in 9 iterations in 1.236 seconds, while the VSC MT-HVDC case takes 12 iterations in 3.703 seconds. The slow-down is mainly due to the increased iterations after the first and only $Vshmin$ violation of VSC 5 on bus 4000.

\begin{table}[]
	\centering
	\caption{Scenario 3: VSC MT-HVDC Results with $Ish_3\le0.19$}
	\label{tab:scenario3_results_VSC}
	\begin{tabular}{@{}llllllll@{}}
		\toprule
		\begin{tabular}[c]{@{}l@{}}Bus\\ $m$\end{tabular} & \begin{tabular}[c]{@{}l@{}}Node \\ $n$\end{tabular} & \begin{tabular}[c]{@{}l@{}}$Psh$\\ {[}pu{]}\end{tabular} & \begin{tabular}[c]{@{}l@{}}$Qsh$\\ {[}pu{]}\end{tabular} & \begin{tabular}[c]{@{}l@{}}$Pdc$\\ {[}pu{]}\end{tabular} & \begin{tabular}[c]{@{}l@{}}$Vm$\\ {[}pu{]}\end{tabular} & \begin{tabular}[c]{@{}l@{}}$Vsh$\\ {[}pu{]}\end{tabular} & \begin{tabular}[c]{@{}l@{}}$Ish$\\ {[}pu{]}\end{tabular} \\ \midrule
		1                                                 & 1                                                   & \cellcolor[HTML]{C0C0C0}0.2                              & 0.6169                                                   & -0.1811                                                 & \cellcolor[HTML]{FFFFFF}1.06                            & \cellcolor[HTML]{C0C0C0}1                                & 0.6118                                                   \\
		3                                                 & 2                                                   & \cellcolor[HTML]{C0C0C0}0.2                              & -0.1188                                                 & -0.2344                                                 & \cellcolor[HTML]{FFFFFF}1.01                            & \cellcolor[HTML]{C0C0C0}1.02                             & 0.2303                                                  \\
		12                                                & 3                                                   & \cellcolor[HTML]{FFFFFF}-0.0105                         & -0.2051                                                 & -0.0718                                                 & \cellcolor[HTML]{FFFFFF}1.0809                          & \cellcolor[HTML]{C0C0C0}1.1                              & \cellcolor[HTML]{C0C0C0}0.19                             \\
		14                                                & 4                                                   & -0.4640                                                 & 0.2093                                                  & 0.4888                                                  & \cellcolor[HTML]{C0C0C0}1.035                           & \cellcolor[HTML]{FFFFFF}1.0203                           & 0.4918                                                  \\ \bottomrule
	\end{tabular}
\end{table}
\section{Conclusions}

In this paper, an automatic control and limit enforcement method for VSC MT-HVDC is elaborated. By using additional equations in the Newton power flow routine, voltage and current limits can be enforced during the iterations by switching out the control equations and substituting in the limits. Case studies verified the proposed method for handling multiple violations during iterations.

Future work involves handling non-convergence in limit violation cases and releasing the package, \textit{Andes}, on GitHub.

\bibliographystyle{IEEEtran}
\bibliography{IEEEabrv,mybib}

\end{document}